\documentclass[preprint,prl,twocolumn,lengthcheck,10pt,bibnotes]{revtex4}
\usepackage{amsfonts}
\usepackage{amsmath}
\usepackage{amssymb}
\usepackage{graphicx}
\begin{document}
\preprint{UATP/04-05}
\title{Dimer Model for Electronic and Molecular Systems and the Intermediate Phase }
\author{F. Semerianov and P. D. Gujrati}
\email{pdg@arjun.physics.uakron.edu}
\affiliation{Department of Physics, Department of Polymer Science, The University of Akron,
Akron, OH 44325, USA}

\begin{abstract}
We introduce a lattice model of dimers with directional interactions as a
paradigm of molecular fluids or strongly correlated Cooper pairs in electronic
systems. The model supports an intermediate phase that is common to both
systems. There are two different ideal glasses having no moblity since they
possess zero entropy. A pairing parameter is introduced to study the
geometrical distribution of holes in various phases.
\end{abstract}
\date{\today}
\pacs{PACS number}
\maketitle

Phillips \cite{Phillips} has provided strong arguments to suggest topological
similarities between the phase diagrams of electronic and molecular systems,
and the presence of generic intermediate phases with novel properties by
considering numerous classical molecular systems and quantum electronic
systems. If true, this will be a major progress in our conceptual
understanding of glasses \cite{Anderson0}. Unfortunately, these topological similarities cannot
be demonstrated because of a lack of reliable analytical solutions. To
overcome this, we propose a phenomenological lattice dimer model (on a regular
lattice) that describes both a molecular and an electronic system. We
investigate its phase diagram by solving it \emph{exactly} on a Husimi cactus
as an approximation of the original lattice. The solution on the cactus goes
beyond the conventional mean-field approximation \cite{PDG}. Dimers represent
the smallest molecules that, in the absence of any hole (collectively
representing hole, solvent, free volume, dopant, impurity or dissociated
dimers), can get into a unique ordered state not because of the regular
lattice structure, but because of physical interactions. On the other hand,
monomeric molecules form an ordered state merely because of the lattice
structure. Thus, dimers can be regarded as the simplest model molecules whose
physics is primarily controlled by interactions and not by the underlying
lattice, although the latter certainly affects it. The extrapolation of the
disordered liquid below the melting temperature $T_{\text{M}}$ gives rise to
the supercooled liquid (SCL), which remains disordered at temperatures where
the crystal (CR) is more stable, and represents the stationary limit of
metastable states (SMS) which can be studied by the use of the partition
function formalism \cite{SGuj}. It is found that the mathematical continuation
of the SCL entropy $S$ becomes \emph{negative} (\emph{entropy crisis})
\cite{Kauzmann} below a positive Kauzmann temperature $T_{\text{K}}$. A
continuous \emph{ideal glass transition} has to be invoked at $T_{\text{K}}$
to avoid the genuine entropy crisis, since states with negative entropy cannot
be\emph{\ observed} in Nature \cite{Gujun,GujC,GujRC}. The SCL freezes at
$T_{\text{K}}$ and turns into an ideal glass (IG) for $T\leq T_{\text{K}}.$
The glass is inert, independent of the temperature, and has zero entropy so
that there can be no mobility in it, since mobility requires thermodynamically
many configurations (positive entropy) to change into each other.

Dimers can also be thought of as representing strongly correlated Cooper pairs
of electrons in high-$T_{\text{c}}$ superconductors \cite{Anderson}, and have
been investigated by various workers \cite{Roksar,Fradkin}. Since the
antiferromagnetic Mott insulator does not support the motion of holes created
on doping, it is thought that there must exist another insulating but
disordered state, the \emph{spin-liquid }\cite{Anderson}, which becomes
superconducting under doping. At higher doping, superconductivity disappears,
which makes the superconductor an \emph{intermediate phase }\cite{Phillips}.
Similarly, it has become gradually apparent in molecular systems (such as
SiO$_{2},$ GeO$_{2}$ that are tetrahedrally bonded) that the short-ranged
orientational bonding plays an important role in giving rise to a
liquid-liquid (L-L) phase transition \cite{Nelson,Roberts} to an intermediate
disordered phase. In ordinary electrolytes, the attractive dipolar
interactions give rise to a similar intermediate phase in the form of
head-to-tail chains \cite{Levesque} Orientational interactions in the dimer
model endow the model with enough richness to predict a transition that can be 
identified as a L-L transition to
an intermediate phase with orientational order that is different from the
crystalline order. The model, when applied to Cooper pairs, also gives rise to
the columnar, staggered, and plaquette phases that have been seen in the
short-ranged resonance-valance bond model (RVB) of Anderson
\cite{Roksar,Fradkin}. The classical dimer model that we consider gives rise
to both an ordered crystal and two disordered ideal glasses at low
temperatures; the latter have higher energies than the $T=0$ crystal.
\begin{figure}
\includegraphics[width=3.4in]{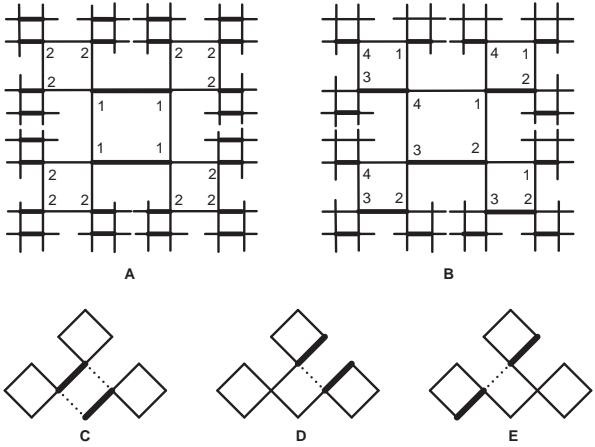}
\caption{Columnar (A) and staggered (B) states on Husimi cactus; the labels
1,2,.. in (A) and 1-4 in (B), respectively, are introduced to capture the
ground states; see description in the text. We show dimers (thick bonds) in
the horizontal directions; they can also point in the vertical directions.
(C),(D) show the resonance due to attractive interaction (dotted line)
$\varepsilon_{\text{p}};$ (E) shows the axial interaction (dotted line)
between colinear dimers.}
\end{figure}

\textbf{Model.} We start with a lattice containing $N$ sites, which we will
take as a square lattice, a bipartite lattice\cite{GujC}. A hole occupies
a lattice site, while a dimer occupies two consecutive sites and the
intervening lattice bond. Let $E$ denote the energy or the energy eigenvalue,
$\mu$ the hole chemical potential, $N_{0}$ the number of the holes, and
$\Omega(N_{0},E)$ the multiplicity of the state. The rigorous thermodynamic
treatment of the model is carried out by using the partition function $Z_{N}$
given by
\begin{equation}
Z_{N}=\sum\Omega(N_{0},E)\eta^{N_{0}}\exp(-\beta E), \label{PF}%
\end{equation}
where $\eta=\exp(\beta\mu)$ is the hole activity$,$ and $\beta=1/T$
($k_{\text{B}}=1)$, $T$ being the temperature. The limit $\eta\rightarrow0$
represents the absence of holes. In the classical treatment, the kinetic term
is absent; the remaining energy $E$ is taken as a sum of various interaction
energies. Excess interaction (energy $\varepsilon)$ between a hole and a dimer
endpoint is sufficient to describe orientation-independent mutual interactions
\cite{Guj}, which we restrict to nearest-neighbor hole-endpoint pairs for
simplicity. There is an orientational interaction (energy $\varepsilon
_{\text{p}})$ between a nearest-neighbor pair of two unbonded dimer endpoints
provided the corresponding dimers are parallel (resonance energy, see Figs.
1C,D). If this interaction is attractive ($\varepsilon_{\text{p}}<0$), the
ground state at $T=0$ is columnar. If the interaction is repulsive
($\varepsilon_{\text{p}}>0$), we introduce an additional \textit{attractive}
axial interaction $\varepsilon_{\text{a}}(<0)$ between the endpoints of two
colinear dimers (see Fig. 1E) so that the ground state at $T=0$ is staggered.
Such ground states also play a central role in the short-ranged RVB model of
high-temperature superconductivity \cite{Roksar}, where the pair of parallel
dimers within a square are said to resonate. Let $N_{\text{p}}$ denote the
number of resonating dimer pairs, $N_{\text{a}}$ the number of
nearest-neighbor axial contacts (endpoint-contacts between colinear dimers),
and $N_{\text{c}}$ the number of nearest-neighbor hole-endpoint contacts, so
that $E=2N_{\text{p}}\varepsilon_{\text{p}}+N_{\text{a}}\varepsilon_{\text{a}%
}+N_{\text{c}}\varepsilon.$ We introduce the Boltzmann weights $w=\exp
(-\beta\varepsilon),w_{\text{p}}=\exp(-\beta\varepsilon_{\text{p}})$ and
$w_{\text{a}}=\exp(-\beta\varepsilon_{\text{a}})$ for later convenience.

The sum in (\ref{PF}) is over $N_{0},N_{\text{p}},N_{\text{a}}$ and
$N_{\text{c}},$ consistent with a fixed $N$. The densities $\phi_{k}$ are the
limiting ratios $N_{k}/N$ as $N\rightarrow\infty;$ here $k=0,$p$,$a, and c. We
always take negative $\mu(<0$) to insure a fully dimer-packed ground state at
absolute zero. The adimensional free energy $\omega=(1/N)\ln Z$ in the limit
$N\rightarrow\infty$ represents the osmotic pressure across a membrane
permeable to the dimers, but not holes \cite{GujState}. The reduced pressure
$\pi_{0}\equiv Pv_{0},$ where $v_{0}$ is the volume of a lattice site, is
given by $\pi_{0}=T\omega-\mu$.\ For $\eta=0,$\ $\pi_{0}$\ is not a useful
quantity. Therefore, it is convenient to introduce the shifted free energy
$\overline{F}(T)=F(T)-F(0),$ which remains meaningful for all $\eta\geq0;$
here $F(T)=-T\omega.$\ The equilibrium state must have the lowest
$\overline{F}(T)$ among all possible states obtained at given $T$,
$w_{\text{p}}$, $w_{\text{a}}$ and $\mu$ \cite{GujState}.\ The 
entropy per site $s$ is given by $s\equiv\omega-\beta e-\phi_{0}\ln\eta.$ At $T=0$, $\phi_{\text{c}}=0$
since there are no holes ($\eta=0$)$.$ The maximum value of $\phi_{\text{p}},$
and $\phi_{\text{a}}$ is 1/2.

The ratio
\begin{equation}
D_{0}\equiv\phi_{00}/\phi_{0} \label{Pair}%
\end{equation}
denotes the number of hole pairs per hole; here, $\phi_{00}$ is the
density of nearest-neighbor contacts between holes. In the RVB model, it is a
measure of how effective is the pairing mechanism for mobile dopants. If all
holes were paired as isolated pairs, we have $D_{0}=1/2.$ It is commonly
believed that holes are always paired in the crystal on a square lattice, but
may not be on other lattices \cite{Dommange}. If holes are separately
clustered as four sites of a square cell, then $D_{0}=1.$ For randomly
distributed holes, $\phi_{00}=2\phi_{0}^{2}$ on the current lattice, and
$D_{0}\equiv2\phi_{0}.$ However, the dimer connectivity will create
correlations even at infinite temperatures, and we expect a higher tendency to
pair so that $D_{0}>2\phi_{0}.$ Our results below confirm the additional
correlations in the disordered phase. It is clear that the values of $\phi
_{0},$ and $D_{0}$\ allow us to draw important conclusions about the way holes
are distributed in the system.

\textbf{Husimi Cactus Solution.} In order to solve the model \emph{exactly},
we replace the square lattice by a site-sharing Husimi cactus, a recursive
lattice, obtained by connecting two squares at each sites. This is
the only approximation we make. The cactus can be thought of as a
checker-board version of the square lattice, representing squares of a given
color \cite{GujC}; the squares of the other color are missing. However, a pair
of dimers on the cactus that would have belonged to a missing square on the
original square lattice is counted as a parallel pair on the cactus
\cite{SGuj}. The model is solved exactly using by now the standard recursive
technique, as discussed elsewhere \cite{PDG,GujC,GujRC}. Because of its
exactness, the calculation (i) respects all local (such as gauge) and global
symmetries, in contrast to the conventional mean-field solution which is known
to violate local symmetries, and (ii) thermodynamics is never violated
\cite{PDG}. The fix-point (FP) solutions of the recursion relations represent
possible states in the system. The free energy is calculated using the method
originally proposed by Gujrati \cite{PDG}. The FP solution that maximizes the
osmotic pressure represents the three stable states that are found: the
ordered phase, i.e. the crystal (CR) at low temperatures, the disordered
phase, i.e. the equilibrium liquid (EL) at high temperatures, and an
intermediate phase (IP) with intermediate orientational order involved in a
liquid-liquid (L-L) transition; see below for a complete description.
\emph{Abandoning} this maximization principle and continuing the FP solutions
allows us to obtain the \emph{stationary} \emph{metastable} states, from which
we construct the \emph{ideal} \emph{glassy} states \cite{Gujun} that replace
the metastable states of negative entropy.

The exact solution can also be taken as the \emph{approximate} theory for the
square lattice. The continuation of EL to lower temperatures describes SCL,
which as we show here exhibits the entropy crisis at $T=T_{\text{K}}$. The
disordered liquid EL undergoes a transition to IP. The continuation of IP to
lower temperatures also exhibits an entropy crisis of its own at a temperature
$T=T_{\text{K}}^{\prime}$ that is usually different from $T_{\text{K}}$. We
study various contact densities in CR, IP and EL, and their continuation.%
\begin{figure}
\includegraphics[width=3.3in]{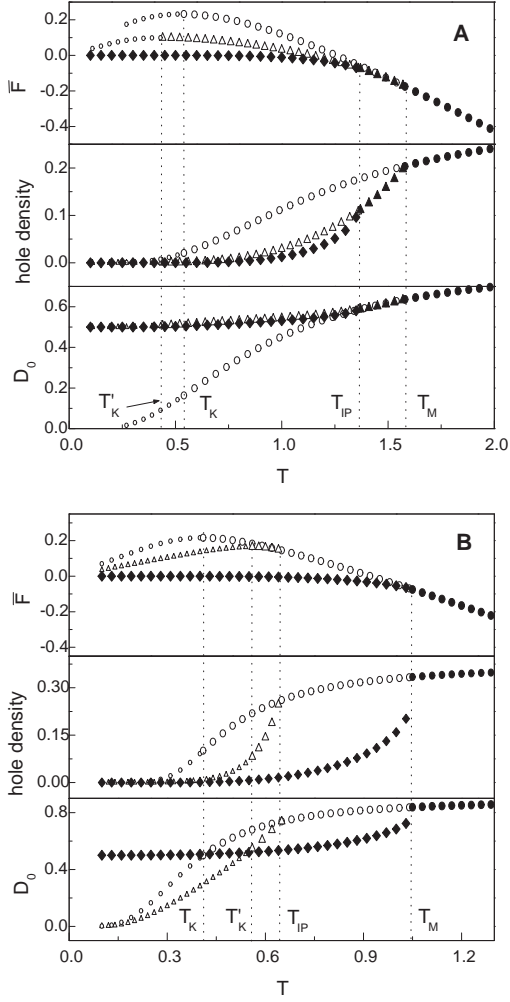}
\caption{(A) $\varepsilon_{\text{p}}=-1,\varepsilon_{\text{a}}=0,$(B)
$\varepsilon_{\text{p}}=1,\varepsilon_{\text{a}}=-1;\ \varepsilon
=0.1,\eta=-0.1$ in both cases. }
\end{figure}

\textbf{Results}. We briefly discuss some important results, with details
found in\cite{SGuj}. For $\eta=0$ $(\mu\rightarrow-\infty)$, we have
analytical results, which clearly establish three distinct phases. For finite
$\mu$, the results are obtained numerically. It should be stressed that the
existence of IP depends on the choice of the energy parameters and the hole
chemical potential. We note that in Fig. 2, the stable states (EL, IP, and CR)
are shown by filled symbols (circles, triangles, and diamonds, respectively),
the realizable metastable states ($S\geq0$) resulting from EL and IP by the
same respective empty symbols, and unrealizable metastable states ($S<0$) by
the smaller empty symbols. The point $S=0$ denotes the Kauzmann point. The
extensions corresponding to the unrealizable states must be \emph{replaced} by
horizontal lines passing through the Kauzmann point (not shown), which
represent ideal glass states that are frozen in their respective metastable
state at Kauzmann point.

\textsc{Disordered Phase}\textbf{. }The disordered phase is characterized by
random orientations of individual dimers, and represents EL at high
temperatures ($T>T_{\text{M}}$) and its analytical continuation SCL at low
temperatures ($T_{\text{M}}>T>T_{\text{K}}$). The disordered phase free
energy, hole density $\phi_{0}$, and $D_{0}$\ are shown in Figs. 2A and 2B 
when the ground state is columnar and staggered, respectively. We immediately
note the existence of the entropy crisis in SCL in both cases below their
respective $T_{\text{K}}$. The states that represent the continuation of SCL
over the temperature range $T<T_{\text{K}},$ where it is unphysical due to
negative entropy must be replaced by an IG at $T_{\text{K}}$ as discussed
above. In terms of $r\equiv\lbrack\lambda/w_{\text{p}}^{2}]^{1/4},$
$\lambda=1+6w_{\text{a}}w_{\text{p}}+(w_{\text{a}}w_{\text{p}})^{2},$ and
$u^{\prime\prime}\equiv2+w_{\text{p}}+w_{\text{a}}+r^{2}w_{\text{p}}^{2}%
,$\ the analytical expressions for thermodynamic quantities for $\eta=0$ are
$\phi_{\text{a}}^{\text{dis}}=(w_{\text{a}}r^{2}+3w_{\text{a}}w_{\text{p}%
}+w_{\text{a}}^{2}w_{\text{p}}^{2})/2r^{2}u^{\prime\prime},$ $\phi_{\text{p}%
}^{\text{dis}}=(w_{\text{p}}r^{2}+3w_{\text{a}}w_{\text{p}}+(r^{4}%
+w_{\text{a}}^{2})w_{\text{p}}^{2})/4r^{2}u^{\prime\prime},$ and
$\omega^{\text{dis}}=(1/2)\ln u^{\prime\prime}-\ln2$. We note that the
high-temperature limit ($w_{\text{a}}\rightarrow1,$ $w_{\text{p}}%
\rightarrow1)$ of $s_{\text{dis}}$ gives $\varphi_{\text{Husimi}}=1.7071$,
where $\varphi=\exp(2s)$ is the average number of configurations per dimer in
the absence of interactions. This is closer to the exact value on the square
lattice, $\varphi_{\text{exact}}=1.7916,$ \cite{KasteleynFisher} than that of 
for Bethe lattice $\varphi_{\text{Bethe}%
}=1.6875$ \cite{Chang}.

\textsc{Ordered \& Intermediate Phases}\textbf{. }The calculation for the
ordered phase requires separate consideration for the two cases. For the
attractive case ($\varepsilon_{\text{p}}<0,\varepsilon_{\text{a}}=0$), shown
in Fig. 2A, the ordered state at $T=0$ is the perfect columnar CR. For this
phase, $T\omega^{\text{ord}}=-\varepsilon_{\text{p}}$, $s^{\text{ord}}=0$,
$\phi_{\text{p}}^{\text{ord}},\phi_{\text{a}}^{\text{ord}}=1/2.$ We take
dimers in the perfect CR to be oriented in the horizontal direction. As $T$ is
raised, CR begins to distort so that $s^{\text{ord}}$ increases, but
$\phi_{\text{p}}^{\text{ord}}$ and $\phi_{\text{a}}^{\text{ord}}$ decrease. As
a result, some dimers begin to resonate in the vertical directions until
resonating dimers in both directions have equal density at $T_{\text{IP}}%
\cong1.37,$ at which point CR undergoes a continuous transition into a new
state IP which is analogous to the plaquette RVB state \cite{Read}. This phase
melts into EL at $T_{\text{M}}\cong1.58$. The continuation of EL below
$T_{\text{M}}$ gives rise to SCL, whose entropy vanishes at $T_{\text{K}}%
\cong0.54.$ Similarly, the continuation of IP below $T_{\text{IP}}$ describes
its metastable state whose entropy also vanishes but at $T_{\text{K}}^{\prime
}\cong0.44$, thus giving rise to another IG. Both IG's have zero entropy, but
have different energies and contact densities, so that they represent two
different glasses. As $\mu$ is decreased further from $\mu=-0.1$, the topology
of the phase diagram does not change; however, the IP entropy now has a weaker
temperature variation and almost no temperature dependence in the limit when
$\mu\rightarrow-\infty$. This has the effect that the melting transition at
$T_{\text{M}}$ becomes first-order. This limit
allows for the analytical treatment, which gives $\phi_{\text{a}}^{\text{IP}%
}=(3w_{\text{p}}+w_{\text{p}}^{2})/2\lambda,$ $\phi_{\text{p}}^{\text{IP}%
}=1/4+\phi_{\text{a}}^{\text{IP}}/2,$ $\omega^{\text{IP}}=(1/2)\ln\left(
\lambda\right)  -\ln\sqrt{2},$ for $w_{\text{a}}=1$. On the other hand,
increase of $\mu(<0)$ shifts both Kauzmann temperatures towards absolute zero,
and IP eventually disappears \cite{SGuj}. Thus, excess holes destroy not only
the IG$^{\prime}$s but also IP; the former was also seen in a polymer model
recently \cite{GujRC}. We note that the IG$^{\text{dis}}$ and IG$^{\text{IP}}$
have almost the same hole density as CR. Despite this, $D_{0}$ for
IG$^{\text{dis}}$ has a much lower value than for IG$^{\text{IP}}$ or CR. It
is clear from $D_{0}\cong0.5$ that most holes in the latter are dimerized. In
the disordered phase, even at high temperatures, $D_{0}>2\phi_{0},$ indicating
that there are correlations induced in the hole distribution due to dimer
connectivity. Loosely speaking, holes in the disordered phase are mostly
isolated, with a tendency to dimerize at lower temperatures, which is evident
from the values of $D_{0}$ near the Kauzmann temperature where $\phi_{0}%
\cong0$; see Fig. 2A.

The staggered ground state for the repulsive case ($\varepsilon_{\text{a}%
}=-1,\varepsilon_{\text{p}}=1$) is completely \emph{frozen }for $\eta=0$
$(\mu=-\infty),$ and has $\phi_{\text{a}}^{\text{ord}}=1/2$, $\phi_{\text{p}%
}^{\text{ord}}=0,$ $s^{\text{ord}}=0.$ It melts at $T_{\text{M}}$ via a
first-order transition into EL. There is a continuous L-L transition at
$T_{\text{IP}}$ in the SCL region where a new IP phase emerges. The latter
corresponds to configurations of dimers preferentially aligned in one
direction (vertical or horizontal). Its SCL continuation below $T_{\text{IP}}$
gives rise to the entropy crisis at $T_{\text{K}}$. The IP also exhibits its
own entropy crisis at $T_{\text{K}}^{\prime}$. The CR state is no longer
frozen for a finite $\mu$ as shown in Fig. 2B; here, $T_{\text{M}}\cong1.05,$
$T_{\text{IP}}\cong0.65,$ $T_{\text{K}}\cong0.41,$ and $T_{\text{K}}^{\prime
}\cong0.56$. For a given $\varepsilon_{\text{p}},$ $T_{\text{IP}}$ moves
towards zero, but $T_{\text{M}}$ decreases to a non-zero value $T_{\text{M}%
}\cong1.3$, as $\left\vert \varepsilon_{\text{a}}\right\vert \rightarrow0$
\cite{SGuj}. On the other hand, rasing $\left\vert \varepsilon_{\text{a}%
}\right\vert $ eventually makes $T_{\text{IP}}>T_{\text{M}}$, hence causing
the L-L transition to appear in the stable region of the phase diagram
\cite{SGuj}. We note that at high temperatures, $D_{0}>2\phi_{0}$ in EL,
indicating the presence of correlations in the hole distribution due to dimer
connectivity. At low temperatures, the holes appear dimerized in CR and the
two IG$^{\prime}$s. While the SCL hole density is larger then that of IP,
$D_{0}^{\text{IP}}<D_{0}^{\text{SCL}}$ indicating higher hole pairing tendency
in IP. We find that $T_{\text{K}}$ as well as $T_{\text{M}}$ decrease and move
to zero as $\mu$ $(<0)$ increases. Adding too many holes moves melting to
$T=0$, thus destroying the CR state and, therefore, the IG transition
\cite{SGuj}.

In conclusion, we have introduced an empirical classical model of dimers, and
presented its exact calculation on a Husimi cactus which serves as an
approximate theory on a square lattice for molecular and RVB systems. The
behavior as a function of hole composition can be studied by varying the
temperature. The calculation provides a strong support for the parallel
between the two systems as observed recently, in particular about the presence
of intermediate phases \cite{Phillips}. The empirical model reproduces all the
known or observed phases in the two systems, which justifies its choice. The
calculation establishes the existence of ideal glasses in the metastable
states, and provides a qualitative comparison of various stable and metastable
states. In particular, attention has been paid to study how holes are
distributed by calculating a pairing parameter $D_{0}$ in various stable and
metastable states, which is another novel feature of this work. Our work also
suggests that the ideal glass related to IP may be a strong candidate for the
spin liquid in the electronic system. We also believe that there is a quantum
analog of IP that could be relevant in high-Tc superconductors.

\end{document}